\documentclass[11pt,hyper]{JHEP3}

\usepackage{amsmath, amssymb, mathrsfs, bbm}

\usepackage[final]{graphicx}
\usepackage{color}
\usepackage{calc}
\usepackage{array}
\usepackage{colortbl}
\usepackage{subfigure}
\usepackage{cite} 
\definecolor{refkey}{rgb}{0.5,0.5,0.5}
\definecolor{labelkey}{rgb}{0.5,0.5,0.5}

\definecolor{lightred}{rgb}{1,0.4,0.4}
\definecolor{lightyellow}{rgb}{1,1,0.5}
\definecolor{lightblue}{rgb}{0.6784,0.8471,0.9020}
\definecolor{lightorange}{rgb}{1,0.75,0.5}
\definecolor{lightgreen}{rgb}{0.8392,0.9234,0.701}
\definecolor{lightviolet}{rgb}{0.8392,0.6236,0.651}

\let\normalcolor\relax

\def\mtilde{\tilde{m}}

\usepackage{bbm}

\newcommand{\beqs}{\begin{equation*}}
\def\beq{\begin{equation}}

\newcommand{\eeqs}{\end{equation*}}
\def\eeq{\end{equation}}

\newcommand{\beqas}{\begin{eqnarray*}}
\newcommand{\beqa}{\begin{eqnarray}}

\newcommand{\eeqas}{\end{eqnarray*}}
\newcommand{\eeqa}{\end{eqnarray}}

\newcommand{\eq}[2]{\begin{equation} #1 \label{#2} \end{equation}}

\newcommand{\eps}{\varepsilon}
\newcommand{\al}{\alpha}
\newcommand{\be}{\beta}
\newcommand{\ga}{\gamma}
\newcommand{\de}{\delta}

\newcommand{\si}{\sigma}

\newcommand{\La}{\Lambda}

\newcommand{\blist}{\begin{itemize}}

\newcommand{\elist}{\end{itemize}}

\providecommand{\href}[2]{#2}

\DeclareFontFamily{OT1}{rsfs}{}
\DeclareFontShape{OT1}{rsfs}{m}{n}{ <-7> rsfs5 <7-10> rsfs7 <10->rsfs10}{} 
\DeclareMathAlphabet{\mycal}{OT1}{rsfs}{m}{n}

\def\yhat{\hat{y}}
\def\Fhat{\hat{F}}
\def\Ahat{\hat{A}}

\def\cL{{\cal L}}
\def\cG{{\cal G}}

\def\cN{{\cal N}}
\def\cM{{\cal M}}

\def\cR{{\cal R}}

\def\cO{{\cal O}}

\def\cF{{\cal F}}

\def\extd{{\rm d}}

\def\tr{{\rm tr}}
\def\Tr{{\rm Tr}}
\def\TrL2{{\rm Tr}_{L^2}}

\def\atbdry{\Big|_{\partial \cM}}
\def\atbdry0{\Big|_{\partial \cM_0}}
\def\atbdry1{\Big|_{\partial \cM_1}}

\newcommand{\bra}[1]{\langle #1 |}
\newcommand{\ket}[1]{| #1 \rangle}

\newcommand{\zbar}{\bar{z}}

\def\N{{\cal N}}

\title{Fayet-Iliopoulos Terms in AdS/CFT with Flavour}
\author{M. Ammon$^a$, J. Erdmenger$^{a,b}$, S. H\"ohne$^a$, D. L\"ust$^{a,b}$ \& R. Meyer$^a$ \\
$^a$Max Planck-Institut f\"ur Physik (Werner Heisenberg-Institut) \\
F\"ohringer Ring 6, 80805 M\"unchen, Germany\\
 \\
$^b$Arnold Sommerfeld Center for Theoretical Physics\\
Department f\"ur Physik, Ludwig-Maximilians-Universit\"at M\"unchen,\\
Theresienstra\ss{}e 37, 80333 Munich, Germany\\
\\
E-mail: \{ammon,jke,hoehne,luest,meyer\}@mppmu.mpg.de}
\abstract{
We construct the gravity dual of a field theory with 
flavour in which there are Fayet-Iliopoulos (FI) terms present. For this purpose we turn on 
a constant Kalb-Ramond B field in four internal space directions of
$AdS_5 \times S^5$ together with a D7 brane probe wrapping $AdS_5 \times
S^3$. The B field induces noncommutativity on the four internal
directions of the D7 brane probe perpendicular to the AdS boundary. We
argue that the moduli space of the Higgs part of mixed Coulomb-Higgs states in the dual
field theory is described by the ADHM equations for
noncommutative instantons on the four internal directions of the D7
brane probe. In particular, the global symmetries match. The FI term
arises as the holographic dual of an anti-selfdual B field.
We discuss possible applications for this construction.}
\preprint{MPP-2008-40, LMU-ASC 27/08}

\begin{document}
\baselineskip 18pt
\parindent 0mm
\parskip 2ex

\section{Introduction and Summary}
\label{sect:intsumm}

In view of making further progress towards generalizing the AdS/CFT
correspondence \cite{Maldacena:1997re,Witten:1998qj,Gubser:1998bc} 
to physically relevant phenomena, it is useful to study
the gravity duals of quantum field theories with non-trivial moduli
spaces. This applies in particular to theories with
flavour added by virtue of probe D7 branes wrapping a subspace of
$AdS_5 \times S^5$ which is asymptotically $AdS_5 \times S^3$
\cite{hep-th/0205236}.  
For instance it has been shown in
\cite{Guralnik:2004ve,Guralnik:2004wq, Guralnik:2005jg,Erdmenger:2005bj}
that the mixed Coulomb-Higgs branch of the $\N=2$ theory with two flavours
is dual to instanton configurations in the supergravity theory. These are
instanton solutions for the $SU(2)$ gauge field in the four directions
of a probe of two D7 branes perpendicular to the AdS boundary.~\footnote{This is based on the fact that Yang-Mills instantons can be
  described by D$p$ branes dissolved inside the worldvolume 
of D$(p+4)$ branes \cite{Witten:1995gx,Douglas:1996uz} - for a review see
\cite{Dorey:2002ik}.} 
These solutions arise since the Dirac-Born-Infeld and Wess-Zumino
contributions to the D7 probe brane action combine in such a way as to
give an action containing only the anti-selfdual part of the field
strength tensor. The instantons are selfdual with respect to the flat
four-dimensional metric. This is due to the fact
that the metric dependence of the 
$D7$ brane action is limited to an overall factor which only depends
on the AdS radial coordinate. Moreover 
it was shown that the Higgs squark vev is dual to the
size of the instanton in the supergravity theory. 
In \cite{Erdmenger:2005bj} the vector meson spectrum of
fluctuations about this instanton background was computed as function 
of the instanton size, and it was shown that the spectra at zero and
infinite instanton size are related by a singular gauge
transformation. In \cite{Arean:2007nh} the instanton analysis was 
extended to all orders in $\alpha'$, with special emphasis on the small
instanton behaviour. 

In this paper we construct the gravity dual of a quantum field theory
in which FI terms are present. For this purpose we consider a
constant Kalb-Ramond $B$ field in the four directions
perpendicular to the AdS boundary, but parallel to the D7 brane probe.
The B field generates non-commutativity of the coordinates 
in these directions \cite{Seiberg:1999vs}. For a D7 brane probe in a
selfdual B field background, the no-force condition is satisfied,
and there is no FI term present. However, for a D7 brane probe in an anti-selfdual B field background a no-force condition does not exist and the supersymmetry breaking is parametrized by an FI term. In this case, the B field is dual to an $\N=2$ FI term, coupling to  the triplet $(D,F_1,F_2)$ of real fields in the dual field theory. The quantum numbers of the B field and the FI term coincide. The analysis of \cite{Billo:2005fg} for the $D3-D(-1)$-system in flat space also supports this claim: The anti-selfdual part of the B field is shown there to induce the FI term in the effective action via string disk diagrams.

In the case of an anti-selfdual B field background, the D3/D7 brane
intersection, which consists of $N_c$ D3 branes generating the $AdS_5
\times S^5$ spacetime in the near-horizon limit and $N_f$ (coincident)
D7 branes, is not static. 
A no-force condition does not exist and the D3 and D7 branes attract each
other. The attractive force is parametrically small when the FI term is small. If the distance of
the D3 and D7 branes is below a critical value, tachyons appear in the
spectrum of 3-7 strings. Consider one of the D3 branes at a distance below the
critical value: After tachyon condensation, it will be dissolved in the D7
brane and can be described by instanton configurations on the D7 brane. The colour direction described by this particular D3 brane will then be broken. If all
D3 branes are dissolved
in the $N_f$ D7 branes, the configuration will be static. We will show that this configuration is supersymmetric, which can also be understood from the gauge theory point of view. However, the probe approximation in AdS/CFT is not applicable for the D7 any longer, if all D3's are dissolved in the D7 brane.

Therefore we  assume in this paper that $k$ D3 branes are dissolved in the
D7 brane with $k\ll N_c$, giving rise to an instanton configuration with charge $k,$ and
that the remaining $N_c-k$ D3 branes are far away from the D7 branes. Such a
configuration is not supersymmetric but is almost stable. Therefore the $N_c-k$ D3 branes will generate the $AdS_5\times
S^5$ metric and the selfdual five-form flux. This approach 
can be viewed as an adiabatic approximation. It
implies that in the probe approximation for D7 branes with charge 
$k$ instanton embedded in $AdS_5 \times S^5$, the instability 
is not visible in the brane embedding, but the tachyonic modes might show up again in the fluctuation spectrum. 

Another indication for the decay of this setup being suppressed is that the $\overline{D3}$-brane charge induced on the $D7$ by the B field can not leave the $D7$ through the on-shell process of condensing first to a zero-size instanton which then detaches from the D7 as a $\overline{D3}$. The reason is that the small instanton limit is no longer available if an FI term is present in the ADHM equations, such that the $\overline{D3}$ would need to go off-shell to condense outside the $D7$. This process might only be possible through a tunneling process, being exponentially suppressed.\footnote{We thank Luca Martucci for this argument.}

We show that although from the geometric point of view supersymmetry is broken
in the presence of an anti-selfdual B-field, the modified ADHM equations for
the noncommutative instantons on the D7 probes can still be identified with
the D- and F-term equations for the Higgs part of mixed Coulomb-Higgs states in
the dual gauge theory on the boundary of AdS. 
In these D- and F-term equations an FI term is present. By Higgs part of a Coulomb-Higgs state we mean the colour directions
for which the squark fields acquire a vacuum expectation value. The other colour directions comprise the Coulomb part of the state.

According to Nekrasov and Schwarz \cite{Nekrasov:1998ss}, 
there is a non-commutative $U(1)$ instanton\footnote{In this paper instantons
  are selfdual. Originally Nekrasov and Schwarz constructed anti-selfdual
  instantons in flat spacetime with selfdual noncommutativity, but call them instantons (see footnote~2~in~\cite{Wimmer:2005bz} to avoid future confusion). In this paper
  we consider selfdual instantons in backgrounds with anti-selfdual B field
  (and therefore anti-selfdual noncommutativity) which can be obtained from
  the original Nekrasov-Schwarz solution by a parity transformation. This
  instanton is also called Nekrasov-Schwarz instanton in this paper.} solution to the
modified ADHM equations. In analogy to the case without B field
\cite{Guralnik:2004ve,Erdmenger:2005bj}, we argue that the flat space
Nekrasov-Schwarz instanton solution remains a solution of the equations of
motion for a D7 probe brane embedded in $AdS_5 \times S^5.$ Note that the
instanton is selfdual with respect to the flat metric in the respective 
directions. Still, it is a solution of the gauge theory on the D7 brane.

The picture which thus emerges is the following: In the presence of an
anti-selfdual B field, a FI term is generated for the gauge theory on
the boundary\footnote{In the context of braneworlds in string theory,
  the generation of FI terms from instantons has been
discussed recently in \cite{Blumenhagen,Haack:2006cy,Akerblom,Kachru,Uranga}. 
Note that 
there, the FI term generation is a quantum effect relying on the
compactification of the internal space, while here in
the AdS/CFT context, the FI term is present already at the classical
level. Note also that in the AdS/CFT context we work with Minkowski signature
and the instanton is located in four directions perpendicular to the AdS
boundary.}. This is supported by the matching quantum numbers for the B field
in the bulk and the auxiliary fields in the boundary field theory. The FI term
is associated with the $U(1)$ factor of the $U(N_c)$ gauge group. 
In fact, the presence of the D7 brane probe is essential for our construction,
since it ensures that the constant background B field may no longer
be gauged away by means of a Ramond-Ramond gauge transformation. 
This implies that the dual field theory at the
boundary has $U(N)$ gauge symmetry instead of $SU(N)$, as is necessary
for an FI term to be present. The $U(1)$ factor corresponds to
singleton degrees of freedom \cite{Aharony:1998qu,Maldacena:2001ss}. 
It will be broken by instanton solutions in the dual gravity theory.

On the gravity side, 
 there are noncommutative $U(1)_F$-instantons on a single D7 brane, which we 
conjecture to be dual to a particular mixed Coulomb-Higgs 
state in the dual gauge theory determined by the instanton charge. These states do not correspond to  supersymmetric vacua of the 
theory, as the D- and F-term equations for the Coulomb part cannot be satisfied in the presence of the FI term. Rather, they are excited states with an excitation energy proportional to the FI parameter. Hence, throughout the paper we refer to them as ``states'' rather than ``vacua'', except in section~\ref{sect:comminst}, where the Coulomb-Higgs vacua are actual true vacua of the theory. 
The size moduli of the instanton configurations are identified with the squark Higgs 
vevs on the gauge theory side. For the selfdual Nekrasov-Schwarz instanton 
on a single D7 brane, we obtain as a nontrivial prediction the existence of 
a new Higgs state in the gauge theory with a single flavour. The Higgs squark 
vev is shown to be given by the square root of the FI term, $q=\sqrt{\zeta}$ 
whereas the squark vev of $\tilde q$ vanishes, $\tilde q=0$.

For the global symmetries, 
we find that an anti-selfdual B-field breaks the 
$SU(2)_L\times SU(2)_R\times U(1)_z$ of the D7 brane configuration 
to $SU(2)_L\times U(1)_R\times U(1)_z$. 
Switching on a selfdual $U(1)_F$ instanton on the D7 brane 
further breaks these symmetries down to 
$SU(2)_L\times {\rm diag} (U(1)_R \times U(1)_F) \times U(1)_z$. 
This corresponds to the symmetries which remain on the field theory side.
The relations $q=\sqrt{\zeta}$, $\tilde q=0$ break the flavor and $U(1)_R$
symmetries to ${\rm diag} (U(1)_R \times U(1)_F)$. Furthermore, these relations break the 
$U(1)$ factor of the gauge group $U(N_c)$.

In the scenario presented, supersymmetry is broken on the Coulomb part of the Coulomb-Higgs branch, but the D and
F term equations (i.e. the ADHM equations) for the Higgs part of the Coulomb-Higgs branch are satisfied simultaneously. It may be possible
to construct gravity duals of theories with metastable vacua
by stabilizing the construction of D3 and D7 branes presented, for
example by placing D7 brane probes in the conifold geometry.
For theories without fundamental flavour degrees of freedom, 
gravity duals of metastable vacua have been discussed for instance in 
\cite{Franco:2005zu,Argurio:2006ew,Argurio:2006ny,Argurio:2007qk,Aganagic:2006ex,DeWolfe:2008zy}. In
these models, the configuration is stabilized by fluxes, 
  leading to configurations of branes at special points, such as the
  resolved conifold. 

Our construction is also similar to models studied within cosmology, 
in the context of inflation. D3/D7 systems with B
field \cite{Dasgupta:2002ew} or on the resolved conifold 
\cite{Krause:2007jk,Baumann:2007ah,Dasgupta:2008hw} have been investigated, and
D term generation has been studied \cite{Haack:2006cy,Dasgupta:2008hw}. 
Our analysis of D3 branes dissolving into D7 branes
bears similarities with the inflationary models of
\cite{Herdeiro:2001zb,Dasgupta:2004dw,Haack:2008yb}.
It will be interesting to generalize the gauge/gravity construction
presented here, in particular by using a suitable stabilization
mechanism, to investigate these relations further.

In AdS/CFT with flavour, the impact of internal B fields on the meson
spectrum has been investigated for the Polchinski-Strassler 
\cite{Polchinski:2000uf} 
and Maldacena-Lunin 
\cite{Lunin:2005jy} backgrounds \cite{Apreda:2006bu,Penati:2007vj}, 
for a review and further references see \cite{Erdmenger:2007cm}. 

This paper is organized as follows: In section \ref{sect:comminst} we review the introduction of flavour degrees of freedom into the holographic setup using probe branes, as well as the description of the Higgs branch of the dual theories using instanton configurations on these probe branes. In sections \ref{sect:BField} and \ref{sect:FI} we show how the noncommutativity induced by the B-field in the internal directions of the probe brane translates into the FI term on the dual gauge theory side. In section \ref{sect:noncomminstantons} we investigate the effect of switching on a noncommutative instanton in this internal noncommutative field theory. We conclude with some discussions of the results in section \ref{sect:conclusions}.
        \section{Commutative Instantons on Flavour Branes}
        \label{sect:comminst}

This section is intended to review instantons on D7 branes which describe the mixed Coulomb-Higgs branch of $\cN=4$ $U(N_c)$ Super-Yang-Mills theory coupled to $N_f$ $\cN=2$ fundamental quark hypermultiplets. In the standard AdS/CFT correspondence, the  $\cN=4$ $U(N_c)$ Super-Yang-Mills theory is realized as the near horizon limit of $N_c$ D3 branes. The type IIB supergravity background is given by 
\begin{eqnarray} \label{eq:background}
ds^2 &=& H_3^{-1/2} \, \eta_{\mu\nu} \, dx^\mu dx^\nu + H_3^{1/2} \left(dy^m dy^m + dz^i dz^i \right),\\ \label{eq:backgroundC4}
C_{(4)}&=&  H_3^{-1} \, dx^0 \wedge \dots \wedge dx^3,\\\label{eq:backgroundgs}
e^\phi &=& g_S = \, \mbox{const.},
\end{eqnarray}
where $r^2=y^m y^m +z^i z^i$ and 
\begin{equation}
H_3(r) = \frac{R^4}{r^4}, \qquad R^4 = 4 \pi g_S N_c \alpha^{\prime 2}.
\end{equation}
Here the Minkowski coordinates are $x^\mu\,,\  \mu=0,\dots,3$, while the internal coordinates are split into two sets, $y^m\,,\  m=4,\dots,7$ and $z^i\,,\  i=8,9$. To couple the dual $U(N_c)$ $\cN=4$ Super-Yang-Mills theory to fundamental matter fields, we follow \cite{hep-th/0205236} and embed $N_f$ $D7$-branes into this background in the way given by table \ref{tab:D-Brane_configuration}. The $D7$-branes fill the $x^\mu$- and $y^m$-directions, while their profile is parametrized by the two transversal directions $z^i(x^\mu,y^m)$. To describe fundamental matter with mass $m$, the embedding of the D7 brane is specified by $z^8=2 \pi \alpha^\prime m$ and $z^9=0.$
\TABLE[!h]{
\arrayrulecolor{white}
        \begin{tabular}{*{10}{c}}
                \multicolumn{10}{>{\columncolor[gray]{.7}}c}
                {Coordinates}\\
                $0$&$1$&$2$&$3$&$4$&$5$&$6$&$7$&$8$&$9$\\ \hline
                \multicolumn{4}{>{\columncolor{lightorange}}c|}
                {$N_c$ D3 branes}& & & & & &\\ \hline
                \multicolumn{8}{>{\columncolor{lightyellow}}c|}
                {$N_f$ D7 branes}& & \\
\multicolumn{4}{
                                                >{\columncolor{lightviolet}}
                                                c|
                                        }
                                        {
                                                $x^{\mu}$
                                        }
                        &
                                \multicolumn{4}{
                                                >{\columncolor{lightviolet}}
                                                c|
                                        }
                                        { \begin{tabular}{cccc}
                                                \multicolumn{4}{c}{$y_m$} \\
                                                \multicolumn{1}{c}{\hspace{-0.7truecm}$\rho$} & \multicolumn{3}{c}{\hspace{0.4truecm}$S^3$}
                                                \end{tabular}
                                        }
                        &
                                \multicolumn{2}{
                                                >{\columncolor{lightviolet}}
                                                c
                                        }
                                        {
                                                $z^i$
                                        }       
                \end{tabular}
\caption{Embedding of the $N_f$ flavour branes.}
\label{tab:D-Brane_configuration}
}
In the stringy picture, i.e. before replacing the D3 branes by their near-horizon geometry, the matter hypermultiplets arise as the massless excitations of strings stretching between the D3 and D7 branes. 
Since we consider $N_f$ to be small, in the limit of large $N_c$ at large (but fixed) 't Hooft coupling $\lambda$ we can ignore the backreaction of the $D7$ branes to the background. This corresponds to the quenched approximation


The dual field theory is an $\mathcal{N}=2$ supersymmetric $U(N_c)$ gauge theory, which has $N_f$ hypermultiplets in the fundamental representation of the gauge group, coupled to the $\cN=4$ vector multiplet in the adjoint representation. 
The scalar components of the latter encode the positions of the D3 branes in the transverse six directions,
\eq{2\pi\al'\Phi_1 = Y^4+iY^5\,,\  2\pi\al'\Phi_2= Y^6+iY^7\,,\  2\pi\al'\Phi_3=Z^8+iZ^9\,.}{eq:transvscalars}
The action, written in $\mathcal{N}=1$ superspace formalism, is given by
\begin{equation}\label{eq:FTBzero}
\begin{split}
\mathcal{L}=& \, \mbox{Im} \,  \int d^4 \theta tr \left( \bar\Phi_I e^V \Phi_I e^{-V} \right) + Q_I^\dagger e^V Q^I + {\tilde Q}^{I\dagger} e^{-V} \tilde Q_I \\ & + \tau \int d^2 \theta \left( tr\left( \mathcal{W}^\alpha \mathcal{W}_\alpha \right) + W \right) + c.c. 
\end{split}
\end{equation}
where the superpotential $W$ is given by
\begin{equation}\label{eq:WBzero}
W= \tr \left( \epsilon_{IJK} \Phi_I \Phi_J \Phi_K \right) + \tilde Q_I \,(m+ \Phi_3)  Q^I
\end{equation}
Here $m$ denotes the mass of the quarks, which we choose to be equal for all hypermultiplets. This theory has a number of global symmetries: For massless quarks, it has a $SO(4,2)\times SU(2) \times SU(2) \times U(1)$ global  symmetry, of which the first factor is the conformal group in four dimensions.\footnote{The $\beta$ function vanishes in the strict $N_c\rightarrow\infty$ limit with $N_f$ small \cite{Kirsch:2005uy}.} In the massive case, the $SO(4,2)$ gets broken to $SO(3,1)$, and the $U(1)$ factor is broken. In both cases, the second $SU(2)$ factor is a $\cN=2$ $\cR$-Symmetry, $SU(2)_R$, while the first $SU(2)$ is an additional global $SU(2)_L$ symmetry, rotating the scalars $\Phi_1$ and $\Phi_2$. 
Additionally, there is a $U(N_f)$ flavour symmetry, the obvious $U(1)$ factor of which  is a  baryon number symmetry \cite{Erdmenger:2007cm}. The different component fields and their quantum numbers are given in table~\ref{tab:quantumnumbers}. We also listed the auxiliary fields $(D,F_1,F_2)$ of the $\cN=2$ $U(N_c)$ vector multiplet $(W_\al,\Phi_3)$, as their $U(1)$ part can couple to a Fayet-Iliopoulos term, which will become important later on. 

The identification of symmetries between the gravity description and the gauge theory is now straightforward: The Lorentz-group $SO(3,1)$ (or the conformal group $SO(4,2)$ if only massless fundamental hypermultiplets are considered) corresponds to the isometries of the induced metric on the embedded D7 brane. 
The internal $SO(4)\simeq SU(2)_L\times SU(2)_R$, which rotates the $y^m$ into each other, is identified with $SU(2)_L\times SU(2)_R$ for a D7 brane. The rotations acting on $\vec{z}$, $U(1)_z$, are identified with the $U(1)$ factor on the field theory side.

\TABLE[ht]{
\arrayrulecolor{black}
\begin{tabular} {|c|c|c|c|c|c|c|c|c|}
\hline
&components & spin & $SU(2)_L \times SU(2)_R$ & $U(1)_{z}$ &$\Delta$ & $U(N_f)$ & $U(1)_F$\\
\hline
$\Phi_1, \Phi_2$ & $X^4, X^5, X^6,  X^7$ &$0$& $(\frac{1}{2},\frac{1}{2})$& $0$ & $1$ & 1 & 0\\
 &$\lambda_1, \lambda_2$ & $\frac 1 2$ & $(\frac{1}{2},0)$& $-1$ & $\frac{3}{2}$ & 1 & 0\\
\hline
$\Phi_3$, $W_\alpha$ & $X_V^A=(X^8, X^9)$ & $0$ &$(0,0)$ & $+ 2$ &$1$ & 1 & 0 \\
&$\lambda_3, \lambda_4$ &$\frac 1 2$& $(0,\frac{1}{2})$& $+1$&$\frac{3}{2}$ & 1 & 0\\
&$v_\mu$& $1$ &$(0,0)$ & $0$& $1$ & 1 & 0\\
&$(D,F_1,F_2)$ & $0$ & $(0,1)$ & $0$ & $2$ & 1 & 0\\
\hline
$Q$, $\tilde Q$ & $q^m=(q, \bar {\tilde q})$&$0$ & $(0,\frac{1}{2})$& $0$&$1$ & $N_f$ & +1\\
&$\psi_i=(\psi, \tilde \psi^\dagger)$ & $\frac{1}{2}$ & $(0,0)$& $\mp 1$& $\frac{3}{2}$ & $N_f$ &+1 \\
\hline
\end{tabular}
\caption{Field content and quantum numbers of the $\cN=2$ theory}
\label{tab:quantumnumbers}
}

The supersymmetric field theory \eqref{eq:FTBzero} with superpotential \eqref{eq:WBzero} has Coulomb- and Higgs vacua, i.e. vacua with nonvanishing expectation value for the adjoint scalars $\Phi_i$, or for the fundamental scalars $q^I$ and $\tilde{q}_I$, respectively. For a mixed choice of color space components of these fields, the corresponding vacua are called mixed Coulomb-Higgs vacua, meaning that some generators of the gauge group are broken down to its respective Cartan subalgebra generators, yielding $U(1)$'s on the Coulomb branch, while other parts of the gauge group are broken completely - this is the Higgs part of the mixed vacuum. 

The vacua are solutions of the F- and D-term equations
\beqa \label{eq:FQBZero}
0 &=& (m + \Phi_3) q^I = \tilde{q}_I (m + \Phi_3) \\ \label{eq:FPhi12BZero}
0 &=& [\Phi_1,\Phi_3] = [\Phi_2,\Phi_3] \\\label{eq:FPhi3BZero}
0 &=& q^I\tilde{q}_I + [\Phi_1,\Phi_2] \\ \label{eq:DBZero}
0 &=& |q^I|^2-|\tilde{q}_I|^2+[\Phi_1,\Phi_1^\dagger]+[\Phi_2,\Phi_2^\dagger]\,.
\eeqa
Here $q,\tilde{q}$ are the squark fields, while $\Phi_i$ are the scalar components of the adjoint transverse scalar superfields, which we  denote by the same symbol as the superfield. The mixed Coulomb-Higgs branch is accessed by solutions of \eqref{eq:FQBZero} with 
\eq{\Phi_3 = \left( 
\begin{array}{cccccc}
\tilde{m}_1& & & & & \\
 &\ddots& & & \\
  & & \tilde{m}_{N-k}& & &\\
   & & &-m& &\\
   & & & &\ddots& \\
   & & & & & -m
   \end{array}
\right)\,,\ q^I = \left( 
\begin{array}{c} 
0\\ \vdots\\0\\ q^I_1\\ \vdots\\ q^I_{k} 
\end{array}
\right)\,,}{eq:coulombhiggs1}
\eq{
\qquad\quad\tilde{q}_I=\left( 0\cdots 0,\tilde{q}^1_I\cdots,\tilde{q}^{k}_{I}
\right)\,.
}{eq:coulombhiggs2}
The remaining equations \eqref{eq:FPhi12BZero}, \eqref{eq:FPhi3BZero} and \eqref{eq:DBZero} then reduce to
\beqa\label{eq:FPhi12BZeroHiggs}
0 &=& (\mtilde_b-\mtilde_a)(\Phi_1)_{ab} = (\mtilde_b-\mtilde_a)(\Phi_2)_{ab}\,,\quad a,b=1,\dots,N-k \\\label{eq:FPhi3BZeroHiggs}
0 &=& q_a^I\tilde{q}_{bI} + [\Phi_1,\Phi_2]_{ab}\,,\quad a,b = 1,\dots,k \\ \label{eq:DBZeroHiggs}
0 &=& |q^I|^2_{ab}-|\tilde{q}_I|^2_{ab}+[\Phi_1,\Phi_1^\dagger]_{ab}+[\Phi_2,\Phi_2^\dagger]_{ab}\,.
\eeqa
Equation \eqref{eq:FPhi12BZeroHiggs} can be solved either by setting the $\mtilde$s equal, or by switching off some off-diagonal elements of $\Phi_{1,2}$ in the corresponding gauge directions. In any case, the first $N-k$ diagonal elements of $\Phi_{1,2}$ may be chosen freely. The off-diagonal $k\times(N-k)$ blocks of $\Phi_{1,2}$ are not constrained by \eqref{eq:FPhi12BZeroHiggs}. The lower  $k\times k$ block is constrained by the last two equations \eqref{eq:FPhi3BZeroHiggs} and \eqref{eq:DBZeroHiggs}. 

With the identification 
\eq{\{{(I)_a}^I,{(J)_J}^b,(B_0)_{ab},(B_1)_{ab}\}\equiv\{q^I_a,\tilde{q}_J^b,(\Phi_1)_{ab},(\Phi_2)_{ab}\}\,,\  a,b=1,\dots,k\,,\  I,J=1,\dots,N_f\,,}{eq:DFADHMIdentCommutative}
equations~\eqref{eq:FPhi3BZeroHiggs} and \eqref{eq:DBZeroHiggs} are just the ADHM equations \cite{Atiyah:1978ri} for the moduli of $k$ gauge instantons in a $SU(N)$ Yang-Mills theory,
\beqa\label{eq:ADHM1}
0 & = & [B_0, B_0^\dagger] + [B_1, B_1^\dagger] + II^\dagger - J^\dagger J \,,\\\label{eq:ADHM2}
0 & = & [B_0, B_1] + IJ\,,
\eeqa
from which all instanton solutions can, in principle, be constructed. We thus just rediscovered the well-known fact \cite{Witten:1995gx,Douglas:1995bn,Douglas:1996uz,Dorey:2002ik} that in a system of $N_c$ $Dp$ and $N_f$ $D(p+4)$ branes, the ADHM equations of $k$ $U(N_f)$ (anti)instantons are just the D- and F-term equations of the intersection $U(N_c)$ theory, parametrizing the mixed Coulomb-Higgs vacua, namely the vacuum in which $k$ generators of the Cartan subalgebra of $SU(N_c)$ are broken (cf.~equation~\eqref{eq:DBZeroHiggs}). Through this identification it is obvious that for only one quark flavour, the theory does not posses Higgs vacua, as there are no nontrivial $U(1)$ instantons in commutative spacetime. This might change, as we will see, if the directions transversal to the $Dp$ branes become noncommutative.

\begin{figure}[t]
\begin{center}
\subfigure[Pure Coulomb branch]{\label{fig:coulombbranch}\includegraphics[height=8cm]{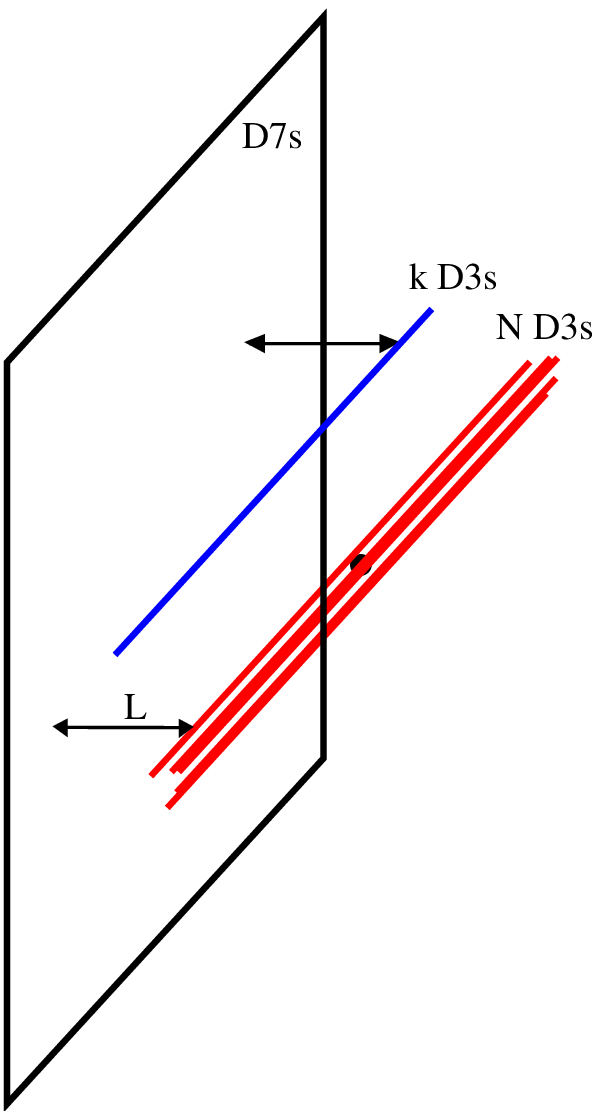}}\hspace{2cm}
\subfigure[Mixed Coulomb-Higgs branch]{\label{fig:higgsbranch}\includegraphics[height=8cm]{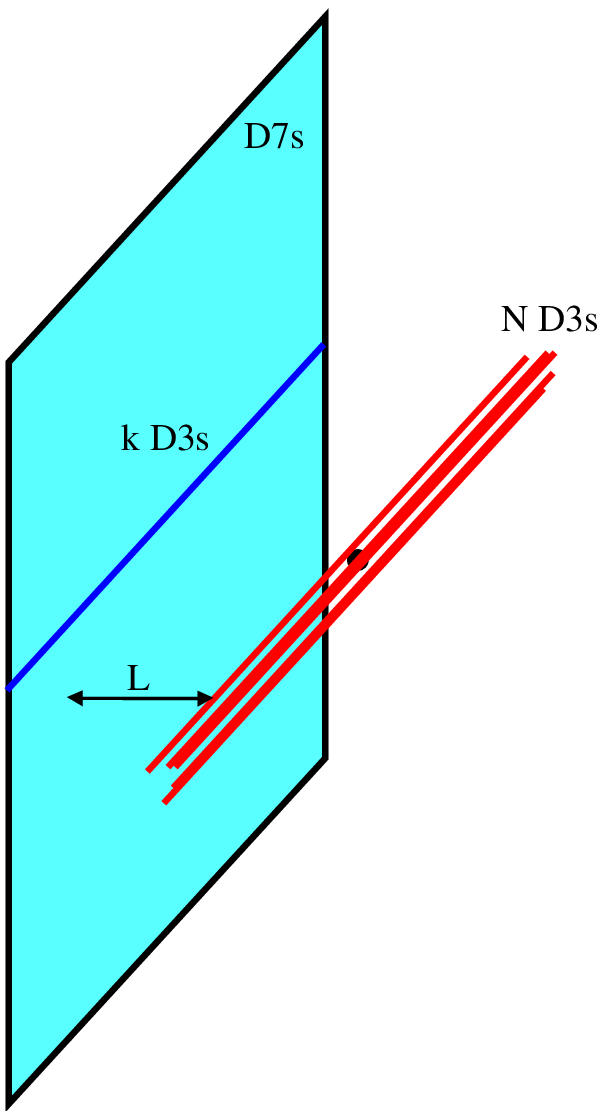}}
\caption{Different supersymmetric vacua in $D3$--$D7$}
\label{fig:D3D7vacua}
\end{center}
\end{figure}


From the point of view of strings and D-branes in flat space, this appearance of instantons as supersymmetric Higgs vacua can be understood as either lower dimensional branes dissolving into the higher dimensional ones \cite{Douglas:1995bn}, or equivalently higher dimensional branes acquiring an induced charge associated with lower Ramond-Ramond p-forms. Both Coulomb and Higgs type breaking of gauge symmetries have D-brane analogues. In figure~\ref{fig:D3D7vacua} the $N_c$ colour $D3$ branes are located at the origin of ten-dimensional Minkowski space, while the $N_f$ flavour $D7$ branes are put parallel to the $D3$ branes, but at a perpendicular separation $|\vec{z}|=L=2\pi\alpha' m$. Coulomb vacua --  cf.~figure~\ref{fig:coulombbranch} -- are configurations with some of the lower-dimensional branes separated from the stack of $N_c$ $Dp$-branes. The transverse scalars \eqref{eq:transvscalars} acquire vacuum expectation values, as they encode the positions of the D3 branes in transverse space. The point in moduli space at which the $k$ separated colour branes coincide with the $N_f$ flavour branes thus also lies on the Coulomb branch. This is only possible as coinciding D3 and D7 branes form a marginal bound state \cite{Witten:1995im}, since some supersymmetry is preserved. 

If some of the D3-branes coincide with the D7 branes, the D3 can dissolve into the D7. This is understood in terms of the low energy effective action on the D7 brane, which includes a coupling to the Ramond-Ramond four form potential $C_4$ via
$$\int\limits_\cM P[C_4]\wedge \Tr(F \wedge F)\,.$$
The Pontryagin density of the flavour gauge field on the D7 brane is a source term for $C_4$, which, upon integration over the transversal coordinates, yields an induced D3 brane charge proportional to the instanton number. Thus field configurations on the D7 stack with nontrivial Pontryagin number in the $\vec{y}$-directions, i.e. instantons, behave as a D3 brane \cite{Douglas:1995bn}, at least for what their charge is concerned. 
Reversely, starting with a dissolved D3 (an instanton) on the D7 brane, one can reach the pure Coulomb branch by sending the size moduli (the vacuum expectation values for the squarks) of the instanton to zero. In this way, the D3 branes sitting on top of the D7 which formerly had a $\Phi_3$ vacuum expectation value $-m$ (cf.~equation~\eqref{eq:coulombhiggs1}), can move away from the D7 brane (changing $\Phi_3$ to some other value). Thus in the zero size limit of the instantons on the D7, the dissolved D3 branes separate themselves from the D7 and can move away from them. The point in moduli space where some or all of the D3s coincide with the D7s thus connects the Coulomb- and Higgs parts of the mixed Coulomb-Higgs branch.


In \cite{Erdmenger:2005bj} it was shown that the Dirac-Born-Infeld and Wess-Zumino parts of the action for $N_f$ D7 branes in $AdS_5\times S^5$ (cf.~equation~\eqref{eq:backgroundC4}) combine to give in order $\al^{\prime 2}$
\beqa\nonumber
S_{D7} &=& \frac{(2\pi\al')^2 T_7}{2}\int\extd^8 \xi \sqrt{g} \Tr (F_{\al\be} F^{\al\be}) - \frac{(2\pi\al')^2 T_7}{2} \int\limits_{\cM_7} P[C_4]\wedge F \wedge F \\\label{eq:noforcecommutative}
&=& \frac{(2\pi\al')^2 T_7}{2}\int \extd^4x \extd^4y \frac{(\vec{y}^2 + m^2)^2}{R^4} F_-^2\,.
\eeqa
Here the (anti)selfdual part of the field strength with respect to the flat metric is defined as
\eq{(F_\pm)_{\al\be} = \frac12\left( F_{\al\be} \pm \frac12 \epsilon_{\al\be\ga\de} F_{\ga\de}\right)\,.}{eq:ASDF}
Equation~\eqref{eq:noforcecommutative} holds for general constant embeddings $|\vec{z}|=2\pi\al'm$, which are the only solutions to the embedding equations of a D7 brane in $AdS_5\times S^5$ consistent with the chosen ansatz \cite{0304032}. Every instanton configuration $F_-=0$ minimizes the D7 action in order $\al^{\prime 2}$ and thus solves the equations of motion for the gauge field on the probe brane. As the instanton-probe brane system has an energy (which is just minus the action) independent of the quark mass $m$ (i.e. the separation from the $N_c$ D3 branes which sourced the $AdS_5\times S^5$ background), it was concluded in \cite{Erdmenger:2005bj} that the instanton-probe brane system in $AdS_5\times S^5$ preserves supersymmetry.
\footnote{In general a static and supersymmetric state of two or more BPS objects (such as D-branes) is again a BPS object, as its mass is  determined only by some charge. The latter is, by charge conservation, determined by the charges of the two constituent BPS objects (see~e.g.~ch.~13.3~of~\cite{Polchinski2}).}
 In \cite{Arean:2007nh}, this argument was pushed further to all orders in $\al'$, which is possible for instantonic configurations, as the radicant of the square root of the DBI action is a complete square, and thus the series expansion in $\al'$ terminates. 

The authors of \cite{Erdmenger:2005bj} then went on in analyzing the symmetries of the instanton-probe brane configuration for two flavours, i.e. $N_f=2$, and the BPST instanton \cite{Belavin:1975fg}, 
\eq{A_\mu=0\,,\  A_m = \frac{2\La^2 \bar{\si}_{nm} y_n}{\vec{y}^2(\vec{y}^2+\La^2)}\  \text{with}\   \bar{\si}_{mn} = \frac{1}{2}\bar{\si}_{\left[m\right.}\si_{\left.n\right]}\,,\  \si_m=(i\vec{\tau},1_{2\times2})\,,\  \bar{\si}_m=\si_m^\dagger\,.}{eq:BPSTInstanton}
They found that the instanton (in singular gauge) breaks $$SU(2)_L \times SU(2)_R \times SU(2)_f\rightarrow SU(2)_L \times \text{diag}(SU(2)_R\times SU(2)_f)$$ by the necessary identification of the space-time $SO(3)$ at infinity with the internal gauge $SU(2)$, which lead them to conclude that the right AdS/CFT identification for this one-instanton solution must be
\eq{q_{i\al} = \frac{\La}{2\pi\al'} \eps_{i\al}\,.}{eq:squarkinstsize}
The $SU(2)_f$ index is $i=1,2$, while the $SU(2)_R$ index is denoted by $\al=1,2$. This is exactly the identification \eqref{eq:DFADHMIdentCommutative} for $k=1$ and $N_f=2$, if one keeps in mind that the squarks transform as a doublet under the $SU(2)_R$ symmetry.

\section{Noncommutativity and Instantons in Flat Space}
\label{sect:BField}

\subsection{B Field Induced Noncommutativity}

Let us first collect some facts about the D3-D7 system with B-field in flat space. Seiberg and Witten showed in \cite{Seiberg:1999vs} that a constant B-field $B_{ij}$ in the Euclidean directions of a flat D-brane embedded in flat spacetime (with metric $g_{ij}$) induces a noncommutative behaviour for the endpoints of open strings ending on the brane through 
\eq{\langle [X^i(\tau,\si=0,\pi),X^j(\tau,\si=0,\pi)] \rangle_{CFT} = i\theta^{ij}\,.}{eq:SWNC}
The noncommutativity parameter is related, in an obvious matrix notation, with the metric and the B-field through
\eq{\theta=2\pi\al'\frac{1}{g+2\pi\al' B}B\frac{1}{g-2\pi\al' B}\,.}{eq:fullthetaflatspace}
Note that because of the additional factor $2\pi\al'$, $B$ has dimension $\text{length}^{-2}$, and thus the dimension in \eqref{eq:SWNC} work out. This result can be obtained without a double scaling limit, but to show that correlators of vertex operators only differ from the zero B-field correlators by the well known noncommutative phase factors \cite{Filk:1996dm}  $e^{-\frac{i}{2} p_i \theta^{ij} k_j}$, the limit $\al'\rightarrow 0$ has to be considered while keeping $\theta$ 
and the open string metric fixed,
\eq{\cG^{-1}=\frac{1}{g+2\pi\al' B}g\frac{1}{g-2\pi\al' B}\,.}{eq:fullGflatspace}
This implies scaling $\al'\sim \sqrt{\eps}$ and $g\sim \eps$, while keeping $B$ fixed.\footnote{Note that this limit corresponds to the infinite 't Hooft coupling limit on the supergravity side, as $\al'\rightarrow 0$ with fixed AdS radius.} In this limit the noncommutativity parameter simplifies to $\theta = B^{-1}$, and  for slowly varying fields the low energy effective theory on the Dp-brane in the directions in which the B-field is nonvanishing reduces  to the well-known noncommutative Yang-Mills theory\footnote{\cite{Szabo:2001kg,Douglas:2001ba} are two concise reviews of the vast topic of noncommutative field theory. Note that for nonabelian theories, the trace runs over the Hilbert space the operators are realized on, as well as over colour space.}  \cite{Seiberg:1999vs}
\eq{S_{\text{NCYM}} = \frac{(\al')^{\frac{3-p}{2}}}{4(2\pi)^{p-2}G_s}\int \Tr \sqrt{\det\cG} \cG^{ij}\cG^{kl} \Fhat_{ik} \ast \Fhat_{jl}\,.}{eq:NCYM}
Here the open string coupling constant is $G_s = g_s \det^{\frac12}((g+2\pi\al' B)g^{-1})$, and the star product is defined as 
\eq{(f\ast g)(x) = f(x)e^{\frac{i}{2}\overleftarrow{\partial^x_i} \theta^{ij} \overrightarrow{\partial^y_j}} g(y)\Big|_{x=y}\,.}{eq:MoyalWeylstar}
The noncommutative gauge field strength
\eq{\Fhat_{ij} = \partial_i \Ahat_j - \partial_j \Ahat_i -i\left(\Ahat_i \ast \Ahat_j - \Ahat_j \ast \Ahat_i \right)}{eq:NCgaugefieldstrength}
also includes a star product. In the $\alpha'\rightarrow 0$ limit, \eqref{eq:NCYM} becomes exact in describing the dynamics of open strings \cite{Seiberg:1999vs}. 

\subsection{Noncommutative U(1) Instantons and modified ADHM Construction}

In this subsection we briefly review the noncommutative ADHM construction, as explained in e.g.~\cite{Chu:2001cx,Furuuchi:2000dx}. 
As shown by Nekrasov and Schwarz \cite{Nekrasov:1998ss}, noncommutative instantons on $\mathbb{R}^4$ can be constructed from the moduli obtained by solving the noncommutative ADHM equations
\begin{eqnarray} \label{eq:NCADHM1}
2(\theta^{45}-\theta^{67}) = \zeta & = & [B_0, B_0^\dagger] + [B_1, B_1^\dagger] + II^\dagger - J^\dagger J \,,\\
0 & = & [B_0, B_1] + IJ
 \,.\label{eq:NCADHM2}
\end{eqnarray}
These equations are very similar in form to the commutative ADHM equations \eqref{eq:ADHM1}-\eqref{eq:ADHM2}, except of the constant term on the left-hand side of \eqref{eq:NCADHM1}. However, the underlying space becomes noncommutative, 
\eq{[\yhat^m,\yhat^n]= i \theta^{mn}\,,\  m,n=4,5,6,7\,,}{eq:NCCoordinates}
where we on purpose denote the transversal coordinates in our D-brane setup \eqref{eq:background} with the same symbol as the coordinate operators. From now on, we omit the hats on operators and leave it to the reader to distinguish between operators and c-numbers where necessary.

As can be seen from \eqref{eq:NCADHM1}, the FI term $\zeta$ appearing in the ADHM construction for instantons (anti-instantons) vanishes for selfdual (anti-selfdual) noncommutativity parameter, i.e. $\theta^{45}=\theta^{67}$, and thus the small instanton singularity in the moduli space is not regulated in these cases. These are exactly the cases in which $\cN=2$ supersymmetry is preserved \cite{Marino:1999af,Dasgupta:2002ew}. We will see later that these are exactly the cases where a no-force condition for the D7 brane in the background of D3 branes holds.

As elaborated in section~\ref{sect:comminst}, without the B-field, the D-term and F-term equations reduced to the Higgs part of a mixed Coulomb-Higgs vacuum of the field theory on the D3-brane are just the ADHM equations for $k$ $U(N_f)$ instantons on the D7 worldvolume transversal to the D3 branes. This statement was checked by direct calculation of the effective action for $D3-D(-1)$ in \cite{Billo:2002hm}, which includes the correct ADHM measure. In \cite{Billo:2005fg} this was extended to the case with B-field on the $D3$ brane. It turned out that the relevant couplings generating the ADHM measure exist also in this case. Furthermore, a disk diagram with $(-1)$ boundary conditions containing an insertion of an auxiliary $D_c$-field and a closed string vertex operator for $B_{\mu\nu}$ generates the correct FI term (cf.~eq.~(6.3)~in~\cite{Billo:2005fg}), 
\eq{\langle V_D V_B \rangle \propto \frac{1}{g_{(-1)}^2} D_c^{-} \bar{\eta}^c_{mn}\theta^{mn}\,.}{eq:DInstantonFITerm}
The Lagrange multiplier $D_c$ then multiplies the bosonic ADHM constraint. As expected, for a configuration of a D-Instanton (and not a  D-antiinstanton) bound to a D3 brane, the FI term only depends on the anti-selfdual part of $B$. The selfdual part gets projected out by the anti-selfdual 't Hooft symbol $\bar{\eta}$. If we now think about the effective theory on the D3-D7 intersection with B-field as connected to the effective action of the D3-D(-1)-System via dimensional reduction, $D_c$ corresponds to the triplet $(D,F_1,F_2)$ of auxiliary fields in the $\cN=2$ vector multiplet, with quantum numbers listed in table~\ref{tab:quantumnumbers}. The index $c$ thus transforms under the $SU(2)_R$-symmetry, and the triplet of fields is in the adjoint representation of the gauge group $U(N_c)$. Furthermore, the gauge coupling on the D-Instanton $g_{(-1)}$  after dimensional oxidation becomes the gauge coupling on the D3-branes of the $D3-D7$ system. Note that the B-field in \cite{Billo:2005fg} is dimensionless, while in this work it has dimension of $\text{energy}^2$. After an appropriate $SU(2)_R$ transformation which sets $(D,F_1,F_2)=(1,0,0)$, one might thus conclude that the low energy effective action on a D3-D7 intersection with a constant B-field in the directions on the D7 brane but transversal to the D3 brane is the same action as without the B-field, but with an additional Fayet-Iliopoulos term
\eq{\frac{\bar{\eta}^3_{mn} \theta^{mn} }{g_{YM}^2} \int\! \extd^4x \,\extd^2 \theta \,\extd^2 \bar{\theta}\, \tr V \,.}{eq:FlatFI}
Note that the normalization is exactly the one to generate the right hand side of equation~\eqref{eq:NCADHM1}, namely  $\bar{\eta}^3_{mn}\theta^{mn} = \zeta$. 

The authors of \cite{Billo:2005fg} also discussed the different possible point particle limits in the $D3-D(-1)$ system. In particular, there is a limit in which $g_{(-1)}$ is held fixed while $\al'\rightarrow 0$, which necessarily decouples the D3 brane degrees of freedom by sending the gauge coupling $g_3$ to zero. Before dimensional reduction, i.e. in the D3-D7 setup, this corresponds to the limit in which the gauge coupling on the $N_c$ D3-branes, $g_{\text{YM}}$, is held fixed, which then implies that 7-7 string degrees of freedom, i.e. the ones on the flavour brane, decouple from the 3-3 and 3-7 strings.
\section{Fayet-Iliopoulos Term in AdS/CFT from Internal Noncommutativity}\label{sect:FI}

We now turn to the analysis of $N_f$ D7 brane probes in the near horizon limit of $N_c$ D3 branes with a B-field switched on. The $N_f$ D7 probe branes are embedded according to table \ref{tab:D-Brane_configuration}. Replacing the $N_c$ D3 branes by the near horizon limit gives the background \eqref{eq:background}-\eqref{eq:backgroundgs}. Additionally a constant Kalb-Ramond B field is switched on in the $\vec{y}$-directions. Due to $H=\extd B = 0$ and a vanishing $C_2$ field,  the supergravity solution \eqref{eq:background}-\eqref{eq:backgroundgs} is not perturbed. In the following we use the skew-diagonalized form
\eq{B = b_1 \extd y^4 \wedge \extd y^5 + b_2 \extd y^6 \wedge \extd y^7\,.}{eq:B-Field}
This form can always be reached by an $SO(4)$ rotation on $\vec{y}$, and may be specialized further to the \textbf{selfdual case} $\mathbf{b_1=b_2=b}$ or to the \textbf{anti-selfdual case} $\mathbf{b_1=-b_2 = b}$. In this background probe D7 branes are embedded by specifying $z^8 (y^m)$ and $z^9 (y^m),$ such that the equations of motion of the D7 brane action $S_{D7}$ are fulfilled,
\begin{eqnarray}
S_{D7} &=& S_{DBI} + S_{WZ}\,, \\ \label{eq:DBIaction}
S_{DBI} &=& - T_7 \int\limits_{D7} \! d^8 \xi \, \text{STr}\sqrt{-\det\left( \mathcal{P}_{ab}[G_{\mu\nu} + B_{\mu\nu}] + 2 \pi \alpha' F_{ab} \right)}\,, \\\label{eq:WDaction}
S_{WZ} &=&  T_7 \int\limits_{D7} \! \text{STr}\,\mathcal{P}[C_{(4)}] \wedge e^{\mathcal{P}[B]+2\pi\alpha'F}\,,
\end{eqnarray} 
where
\begin{equation}
T_7 = \frac{1}{g_S \, (2 \pi)^{7} \alpha^{\prime 4}}\,.
\end{equation}
The pull-back of the background metric $g_{\mu\nu}$ and Kalb-Ramond-field $B_{\mu\nu}$ is denoted by $\mathcal{P}.$ $\text{STr}$ is the symmetrized trace.

One can show that in the presence of a B field the usual embedding $z^8=2 \pi \alpha^\prime m$, $z^9=0$ is no longer a solution of the equations of motion of the D7 brane action, unless $m=0$. Therefore we only study massless embeddings of the D7 probe branes, i.e. 
\eq{z^8=z^9=0\,.}{eq:masslessemb}

Due to the analysis of the flat space D3-D7 system of section 3.2, we expect that the FI term \eqref{eq:FlatFI} survives the decoupling limit of AdS/CFT and thus a description of the mixed Coulomb-Higgs branch of the dual field theory in terms of noncommutative instantons on probe branes is possible, along the lines of \cite{Guralnik:2004ve,Guralnik:2004wq,Guralnik:2005jg,Erdmenger:2005bj,Apreda:2005yz,Apreda:2005hj,Apreda:2006ie,Arean:2007nh}. This is supported by the fact that the deformation induced on the instanton moduli space by the FI term is rather mild, as only the small instanton limit is affected. We therefore expect the map between instantons on the probe branes and the Coulomb-Higgs branch on the field theory  is still possible, in analogy to the case without B-field \cite{Guralnik:2004ve,Erdmenger:2005bj}. 

\subsection{Global Symmetries}

Let us first analyse the geometric symmetries of a D7 brane with the B field \eqref{eq:B-Field}. 
As can be seen from \eqref{eq:background}, the $(\vec{y},\vec{z})$-directions perpendicular to the boundary of $AdS_5\times S^5$ are flat up to the warp factor. 
The components of the B field can then be written as an antisymmetric $6\times 6$-matrix $B_{ij}=-B_{ji}$, transforming in the antisymmetric tensor representation $[1,0,1]=\mathbf{15}$ of the six-dimensional rotation group $SO(6)$.  To find the possible representations compatible with the symmetry breaking induced by the probe brane, we decompose this representation of $SO(6)$ into irreducible representations of the symmetry group which is preserved by the flavour D7 brane, $SU(2)_L\times SU(2)_R \times U(1)_z$. The branching rule for $\mathbf{15}$ of $SO(6)$ into $SU(2)_L\times SU(2)_R \times U(1)_z$ is, according to table 58 of \cite{Slansky:1981yr},
\eq{\mathbf{15} = (0,0)_0 \oplus (1,0)_0 \oplus (0,1)_0 \oplus (\frac12,\frac12)_2 \oplus (\frac12,\frac12)_{-2}\,.}{eq:15decompose}
Since we switch on a B field in the $\vec{y}$-directions only, the $U(1)_z$ charge is zero, which leaves only the first three terms on the right hand side of \eqref{eq:15decompose}. The first term corresponds to a B field in the $\vec{z}$-directions. The second and third terms in the decomposition are, respectively, the selfdual and anti-selfdual parts of the B field in the $\vec{y}$ directions  (cf.~e.g.~\cite{PvNInstantons}). These are the field configurations we are interested in. These B field contributions  are dual to scalar operators in the dual field theory. 

A general B field in the $\vec{y}$-directions breaks the $SU(2)_L\times SU(2)_R\simeq SO(4)$ rotation invariance down to $U(1)_L \times U(1)_R$. The selfdual case $b_1=b_2=b$ preserves the $SU(2)_R$, while the anti-selfdual case preserves the $SU(2)_L$. In the case of a selfdual B field, there is no FI term in the dual gauge theory. However, for an anti-selfdual B field transforming in the $(0,1)_0$ of the $SU(2)_L\times SU(2)_R \times U(1)_z$, there is a FI term present: The anti-selfdual B field in the representation $(0,1)_0$ of $SU(2)_L\times SU(2)_R \times U(1)_z$ has the right quantum numbers to couple to the auxiliary field triplet $(D,F_1,F_2)$, which transforms under $(0,1)_0$ of $SU(2)_L\times SU(2)_R \times U(1)$ (see section~\ref{sect:comminst}). This is consistent both with the brane picture, in which the FI term \eqref{eq:FlatFI} in the D3-D(-1) system only depends on the anti-selfdual part of the B field, as well as with the analysis of the noncommutative ADHM construction \cite{Chu:2001cx}. For instantons, the small instanton singularity is only resolved if the noncommutativity parameter is not purely selfdual, and vice-versa for anti-instantons. The global symmetries on the gravity side are thus consistent with the existence of a holographic coupling of the form \eqref{eq:FlatFI}, in the standard holographic sense, of the B field to the auxiliary field triplet $(D,F_1,F_2)$.

\subsection{Scaling Dimensions}

Also the scaling dimensions work out for such a holographic coupling: The
Fayet-Iliopoulos term has scaling dimension $\Delta=2$. This can also be
deduced as follows. The chiral primary operator of the N=2 gauge multiplet
is given by the scalar component of $\Phi_3$ and hence has scaling
dimension 1, which is determined by the $\cR$ symmetry. By applying twice the 
supersymmetry generators on the chiral primary we obtain the FI-term triplet
$(D, F_1, F_2)$. Therefore the scaling dimension of the FI terms is $\Delta =
1 + 2 \cdot \frac 1 2 =2.$

The quantum numbers of the $\cN=2$ vector multiplet, as can be seen from
the second line in table~\ref{tab:quantumnumbers}, imply that the auxiliary
fields $(D,F_1, F_2)$ transform in the $(0,1)_0$ representation of
$SU(2)_L\times SU(2)_R  \times U(1)_z$, with scaling dimension $\Delta=2$
and zero $U(1)_F$ charge. Thus the coupling \eqref{eq:FlatFI} has the right
scaling dimensions, as well as the correct quantum numbers under the global symmetries, to be interpreted as the holographic version of the coupling \eqref{eq:FlatFI} of the noncommutativity parameter to the triplet $(D,F_1, F_2)$.

\subsection{Supersymmetry and No-Force Conditions}

Another check is that the supersymmetry breaking pattern derived from the $Dp-D(p+4)$-system in flat space \cite{Seiberg:1999vs,Billo:2005fg} coincides with the pattern derived from no-force conditions of probe branes in $AdS_5\times S^5$. In flat space, a $Dp-D(p+4)$ system is $\cN=2$ supersymmetric exactly if $B$ is selfdual. The lower-dimensional brane can then be viewed as an instanton in the four additional directions of the $D(p+4)$-brane. Equivalently, $\overline{Dp}-D(p+4)$ or $Dp-\overline{D(p+4)}$ are supersymmetric if $B$ is anti-selfdual. This configuration corresponds to an anti-instanton. In the cases where $B$ has the selfduality properties which do not lead to a no-force condition, a Fayet-Iliopoulos term is expected to be generated. 

To calculate the unbroken supersymmetry of the $D3-D7$ system we embed a D7 probe brane into the background given by \eqref{eq:background}-\eqref{eq:backgroundgs} and \eqref{eq:B-Field}. Following \cite{Dasgupta:2002ew} we present the $\kappa$-symmetry calculation for one probe D7 brane. We choose $(x^\mu, y^m)$ as our set of worldvolume coordinates and consider only massless embeddings, i.e. $z^8=z^9=0.$ To describe dissolved D3 branes in the D7 probe brane, a $U(1)$ field strength $F_{ab}$ is switched on on the worldvolume of the D7 brane in the directions $y^m.$ This D7 probe brane preserves some supersymmetries if there are non-trivial spinor solutions to the equation \cite{Cederwall:1996ri}
\eq{ \Gamma_\kappa \epsilon = \epsilon \,,}{eq:kappa}
where $\epsilon$ is a Killing spinor and the $\kappa$-symmetry projector $\Gamma_\kappa$ for a D7 probe brane in this background is given by \cite{Bergshoeff:1997kr}
\eq{\Gamma_\kappa = e^{-a} \left(i \sigma_2\right) \otimes \Gamma_{01234567}\,,}{eq:kappaprojection}
where $a$ is a function of $Y_{ik},$ which depends on $\mathcal F_{ij} = \mathcal{P}[B]_{ij} + 2 \pi \alpha^\prime F_{ij}$ in a nonlinear way,
\eq{a=\frac 1 {2 \sqrt{H_3}} Y_{jk} \sigma_3 \otimes \Gamma^{jk}\,.}{eq:kappaa}
Using $\Gamma^{ij} \Gamma_{4567}=-\frac 1 2 \epsilon^{ijkl} \Gamma_{kl}$ for $i,j,k,l$ from $4$ to $7,$ one can rewrite $\Gamma_\kappa \epsilon = \epsilon$ in the form
\eq{\exp\left(-\frac 1 {4 \sqrt{H_3}} \, \sigma_3 \otimes \Gamma^{ik} \left[ Y_{ik}^+ \left(1-\Gamma_{4567}\right) + Y_{ik}^- \left(1+\Gamma_{4567}\right)\right]\right) \left( i \sigma_2 \right) \otimes \Gamma_{01234567} \epsilon = \epsilon}{eq:kapparewritten}
with $Y_{ik}^\pm = \frac 1 2 \left( Y_{ik} \pm (\star Y)_{ik} \right).$ The dual two-form $\star Y$ is calculated with respect to the flat metric. 

Without loss of generality we consider selfdual field strengths on the D7 brane. This implies that only $F_{ij}^+$ (and therefore $Y_{ij}^+$) will depend on the worldvolume coordinates of the $D7$ brane. The kappa symmetry projection cannot be solved since $\epsilon$ has to be a constant spinor. We therefore have to demand $\left(1-\Gamma_{4567}\right) \epsilon = \epsilon.$ 

For a selfdual B field, $\mathcal{F}_{ij}$ and consequently $Y_{ij}$ are selfdual, i.e. $Y_{ik}^-=0.$ Supplemented by $(i \sigma_2) \otimes \Gamma_{0123} \epsilon = \epsilon,$ which is fulfilled by the background, we see that $\Gamma_\kappa \epsilon = \epsilon$ is indeed fulfilled. Since without B field and without instanton we obtain the same conditions for a probe D7 brane, $\cN=2$ (Poincare) supersymmetry is preserved.

In the case of an anti-selfdual B field $Y_{ik}^-$ is no longer zero. Due to the factor $H_3^{-1/2}$ in the exponent of \eqref{eq:kapparewritten}, which also depends on the worldvolume coordinates of the $D7$ brane, we additionally have to satisfy $\left(1+\Gamma_{4567}\right) \epsilon = \epsilon.$ Thus the spinor $\epsilon$ has to satisfy both conditions $\left(1\pm\Gamma_{4567}\right) \epsilon = \epsilon.$ Therefore supersymmetry is completely broken, which is in agreement with field-theoretical considerations. Note that in globally supersymmetric theories a partial breaking of supersymmetry (i.e. from $\cN=2$ to $\cN=1$) is not possible \cite{Witten:1981nf}.

This supersymmetry pattern for instantons can be confirmed by no-force conditions for $N_f$ D7 probe branes. 
An expansion of \eqref{eq:DBIaction} up to $\cO(\al^{\prime 2})$ yields for selfdual $B$ ($b_1=b_2=b$)
\begin{equation}
\label{eq:effective_action_FI}
S_{DBI} + S_{CS} = -T_7 \int \!\! \mathrm{d}^4 x \int \!\! \mathrm{d}^4 y \, \, \left(1+\tfrac{1}{2} \frac{\left( 2\pi\alpha'\right)^2}{H_3+b^2}\tr F_-F_-\right)\,,
\end{equation}
where $F_-$ is defined in \eqref{eq:ASDF}. The trace runs over $U(N_f)$ indices. For calculational details see  Appendix~\ref{sect:determinant_expansion}. The crucial feature in the calculation is that contributions of $\cO(\al')$ cancel between the DBI-part and the Wess-Zumino part of the brane action.

Thus, despite the presence of the B field, the conclusion from this calculation is as in the case without B field, described in  section~\ref{sect:comminst}: There is no force exerted on probe branes if both the B field and the field strength on the branes are selfdual, and instantonic solutions still solve the gauge theory living on the D7 branes, at least to $\cO(\al^{\prime 2})$. For an anti-D7-brane in an anti-selfdual $B$ field one would find that anti-instantons are force-free.

Assuming that the full DBI action \eqref{eq:DBIaction} holds for instantonic configurations, the no-force condition \eqref{eq:effective_action_FI} can also be obtained in a different way, following \cite{Arean:2007nh}: Rewrite the DBI action as 
\begin{align}
S_{DBI} &= - T_7 \int \extd^4 x \extd^4 y \sqrt{-\det\left(P[G] + \cF \right)} \\
&= - T_7 \int \extd^4 x \extd^4 y \sqrt{-\det\left(\eta_{\mu\nu}/\sqrt{H_3}\right)} \sqrt{\det\left(\sqrt{H_3} \delta_{ij} + \cF_{ij} \right)}\\
&= - T_7 \int \extd^4 x \extd^4 y \sqrt{\det\left(\delta_{ij} + M_{ij} \right)}\,.
\end{align}
Here we defined the Levi-Civita symbol $\tilde{\epsilon}_{01234567} = +1$ and the volume element through $\extd^4x \extd^4 y = \frac{1}{8!} \tilde{\epsilon}_{\mu_1\dots\mu_8} \extd x^{\mu_1} \wedge\dots\wedge \extd x^{\mu_8}$. Now one can use the identity for antisymmetric 4 by 4 matrices M with definite duality properties $(*M)_{ij} = \frac12 \epsilon_{ijkl} M_{kl} = \pm M_{ij}$,
\begin{equation}
\det(\mathbf{1} + M) = 1 + \frac12 M^2 + \frac{1}{16}(\ast M M)^2 = \left( 1+\frac{M^2}{4}\right)^2\,.
\end{equation}
Here $M^2 = M_{ij} M_{ij}$. Thus for (anti)selfdual $\cF=B+2\pi\al' F$,
\begin{equation}
\ast(B+2\pi\al' F)=\pm (B+2\pi\al' F)\,,
\end{equation}
the DBI action acquires a simpler square-root free form,
\begin{equation}
S_{DBI} = -T_7 \int \extd^4 x \extd^4 y \left(1+\frac{M^2}{4}\right)\,.
\end{equation}
The Chern-Simons action can be evaluated in a similar way, yielding
\begin{equation}
S_{CS} = \pm T_7 \int \extd^4 x \extd^4 y \frac{M^2}{4}\,,
\end{equation}
where the sign depends on the duality properties of $M$. Thus the D7 brane action reads
\begin{equation}
S_{D7} = S_{DBI} + S_{CS} = -T_7 \int \extd^4 x \extd^4 y \left(1 + \frac{M^2}{4}\left(1 \mp 1 \right)\right)\,,
\end{equation}
Therefore $B+2\pi\al' F$ has to be selfdual for a D7 brane in order to have a force-free situation. For an anti D7 brane, the opposite holds. If $B$ and $F$ have different selfduality properties, supersymmetry will be broken and an FI-term will be created.  Table~\ref{tab:noforce} lists the different possibilities for (non-)supersymmetric configurations.

\TABLE[h]{
                \begin{tabular}{|c|c|c|}\hline
                        \multicolumn{1}{>{\columncolor[gray]{0.95}}c}{} & \multicolumn{1}{>{\columncolor{lightviolet}}c}{$\ast B = +B$} & \multicolumn{1}{>{\columncolor{lightviolet}}c}{$\ast B = -B$} \\\hline
                        \multicolumn{1}{>{\columncolor{lightblue}}c}{D3-D7} & \multicolumn{1}{>{\columncolor{lightorange}}c}{SUSY if $\ast F = + F$} & \multicolumn{1}{>{\columncolor{lightyellow}}c}{no SUSY, FI} \\\hline
                        \multicolumn{1}{>{\columncolor{lightblue}}c}{D3-$\overline{D7}$} & \multicolumn{1}{>{\columncolor{lightyellow}}c}{no SUSY, FI} & \multicolumn{1}{>{\columncolor{lightorange}}c}{SUSY if $\ast F = - F$} \\\hline
                \end{tabular}
        \caption{Supersymmetry conditions}
        \label{tab:noforce}
}

\subsection{Duality Conjecture}\label{sect:dualityconjecture}

We showed above that the global symmetries of the gravity background with B field, the scaling dimensions of the involved fields as well as the supersymmetry breaking pattern are consistent with the generation of an FI coupling of the form \eqref{eq:FlatFI} in the dual field theory. We thus conjecture:
\begin{quote}
The degrees of freedom of $N_f$ D7 brane probes with $k$ units of instanton charge on its worldvolume, embedded into $AdS_5\times S^5$ with a constant B field in the $\vec{y}$-directions, are dual in the standard holographic sense to mixed Coulomb-Higgs states of the dual field theory at strong coupling. The mixed Coulomb-Higgs states of this field theory are described by \eqref{eq:FQBZero}-\eqref{eq:FPhi3BZero} with $m=0$. The field theory D-term equation  \eqref{eq:DBZero} is modified by  an FI term, 
\eq{\zeta\mathbbm{1}_{N_c\times N_c} = |q^I|^2-|\tilde{q}_I|^2+[\Phi_1,\Phi_1^\dagger]+[\Phi_2,\Phi_2^\dagger]\,.}{eq:DFIHiggs}
\end{quote}
In the next section we study noncommutative instantons satisfying the noncommutative ADHM equations on the supergravity side. An explicit example is the Nekrasov-Schwarz instanton for a $U(1)$ gauge theory \cite{Nekrasov:1998ss}. The ADHM equation for the Nekrasov-Schwarz instanton coincide with the D- and F-term equations of the Higgs part of the Coulomb-Higgs branch. This are precisely the directions in which the squarks acquire a vev. The question of consistency of this approach was discussed in the introduction. 

Note that in the AdS/CFT correspondence, a gravity dual does not only describe a dual field theory, but also fixes a particular state (which does not necessarily need to be a vacuum state), in which relevant quantities are computed \cite{Skenderis:2008dh}. Our proposal here thus is that the D7 probe with instanton charge $k$ describes nonsupersymmetric Coulomb-Higgs states, which are, however, not vacua of the theory.

\section{Field Theory Implications of Noncommutative Instantons}
        \label{sect:noncomminstantons}

In this section we study some predictions of the proposed correspondence in the explicit example of a noncommutative $U(1)$ instanton. We consider $U(1)$ instantons, as they are do not have a commutative counterpart and thus should give rise to new phenomena in the dual gauge theory which may be easily identified.

\subsection{U(1) Nekrasov-Schwarz Instanton}

In this section the instanton constructed by Nekrasov and Schwarz is reviewed. Following \cite{Nekrasov:1998ss} we introduce complex coordinates $z_0 = y^4 + i y^5$, $z_1 = y^6+iy^7$, with commutators 
\eq{[z_0,\zbar_0]= 2\theta^{45}\,,\  [z_1,\zbar_1]= 2\theta^{67}\,,\  [z_0,z_1]=0\,.}{eq:NCComplexCoordinates}
Here $\theta^{mn}$ is given in a skew-diagonalized form similar to \eqref{eq:B-Field}. This can be achieved by an appropriate $SO(4)$ rotation. This basis has the advantage that the complex coordinate operators can be mapped to two copies of standard bosonic Fock space generators. For $\theta^{45}$ and $\theta^{67}$ both negative,\footnote{By this choice, we follow the conventions of \cite{Nekrasov:1998ss}.} the map is $a = \zbar_0/\sqrt{2|\theta^{45}|} = (a^\dagger)^\dagger$, $b = \zbar_1/\sqrt{2|\theta^{67}|} = (b^\dagger)^\dagger$. If one or both of the components of $\theta$ change sign, the roles of creation and annihilation operators are interchanged. Again, $\theta^{45}=\pm\theta^{67}$ respectively correspond to a purely selfdual or anti-selfdual noncommutative space. In this section we consider an anti-selfdual $\theta$.

The solution constructed in \cite{Nekrasov:1998ss} is a anti-selfdual one, i.e. an anti-instanton. Within the framework of the D3-D7 system with an Fayet-Iliopoulos term, we need to consider an instanton instead. By a parity transformation $y^5\mapsto -y^5$, the solution of \cite{Nekrasov:1998ss} is straightforwardly changed to be selfdual, 
\beqa 
A & = & \frac{1}{d\left(d+\frac{\zeta}{2}\right)} \left[ z_0 \extd \zbar_0 + \zbar_1 \extd z_1\right]\,, \label{eq:oneinstantonA}\\
F &=& \frac{\zeta}{(d-\zeta/2)d(d+\zeta/2)}\left[f_3(\extd z_0 \extd \zbar_0 + \extd z_1 \extd \zbar_1) + f_+ \extd z_0 \extd z_1 + f_- \extd \zbar_1 \extd \zbar_0\right]\,,\label{eq:oneinstanton}
\eeqa
with $f_3 =   z_1 \zbar_1-\zbar_0 z_0$, $f_+ = 2 \zbar_0 \zbar_1$ and $f_- = 2 z_1 z_0$. Since a parity transformation on euclidean $\mathbf{R}^4$ exchanges selfduality with anti-selfduality, this solution is obtained by exchanging $y^5\mapsto - y^5$, which exchanges $z_0$ with its adjoint. Although $d= \sum\limits_{i=4}^7 (y^i)^2$ is invariant under parity, its form in complex coordinates changes, compared to \cite{Nekrasov:1998ss}, to $d=\zbar_0 z_0 + \zbar_1 z_1$. Note that the parameter $\zeta = \theta^{45}/4=-\theta^{67}/4$ is negative in our conventions (which are those of \cite{Nekrasov:1998ss}). Clearly equation~\eqref{eq:oneinstanton} fulfills the complexified selfduality equations
\eq{F_{z_0\zbar_0}=F_{z_1\zbar_1}\,,\quad F_{\zbar_0z_1}=F_{z_0\zbar_1}=0\,.}{eq:complexselfduality}
It is easy to show that this solution has instanton number plus one by first showing that anti-selfduality $F_+=0$ changes into selfduality $F_-=0$ under parity, and then realizing that the Lagrange density given in \cite{Nekrasov:1998ss},
\begin{equation} \label{eq:NCU1Lagrangian}
\cL = -\frac{1}{8\pi^2} F_{mn}F_{mn} = \frac{\zeta^2}{4\pi^2} \frac{ 1 }{ d^2 \, ( d-\frac \zeta 2) \, ( d+\frac \zeta 2)} \, \Pi\,,
\end{equation}
is invariant under $z_0\mapsto\zbar_0$. The reason is that this only amounts to exchanging the annihilators and creators in the $z_0$ Fock space. The projector $\Pi = \mathbbm{1} - \ket{0,0}\bra{0,0} = \mathbbm{1} - :e^{-a^\dagger a - b^\dagger b}:$ is normal ordered \cite{Furuuchi:2000vc} and thus invariant. The integral-trace  over the Lagrange density stays positive under parity, while the relation between the latter and the instanton number changes sign, and thus the solution \eqref{eq:oneinstanton} needs to have Pontryagin number $+1$, as the anti-instanton of \cite{Nekrasov:1998ss} has $-1$.

The one-instanton solution \eqref{eq:oneinstanton} does not have a freely selectable size modulus, in contrary to the BPST instanton \cite{Belavin:1975fg}. This is expected since the dimensionality of the instanton moduli space is $4 N_f k$. The $N_f=1$ one-instanton solution thus has a four-dimensional moduli space, which is a copy of $\mathbb{R}^4$ encoding the instanton position, but it has no size modulus. Note however that a size of the instanton can still be defined by the squark VEV (cf.~equation~\eqref{eq:k1Nf1}) $q = \sqrt{\zeta} = \sqrt{2|\theta^{45}-\theta^{67}|}$. It is fixed by the FI term. This also is exactly the minimal separation above which a $Dp-D(p+4)$-system in flat space does not have a tachyon \cite{Dasgupta:2002ew}. Exactly at this separation one of the 3-7 string modes becomes tachyonic, and the system ends up in the state with an instanton on $D(p+4)$ after tachyon condensation~\cite{Wimmer:2005bz}.  The squark VEV sets the instanton size as expected, as we argued in section~\ref{sect:BField} for the equivalence between D- and F-term equations on the Higgs branch and ADHM equations also in the noncommutative setup.

The one-instanton background further breaks the remaining symmetries. Using as a basis of rotations in $\mathbb{R}^4$ the rotations in the plane of two directions, e.g. the $4-5$-plane, one can parametrize an infinitesimal $SU(2)_L\times U(1)_R$ rotation of the coordinate operators by 
\beqa\label{eq:trafo1}
z_0' &=& (1+i(c+d)) z_0 + (a+ib) \zbar_1 \\\label{eq:trafo2}
\zbar_0' &=& (1-i(c+d)) \zbar_0 + (a-ib) z_1 \\\label{eq:trafo3}
z_1' &=& (1+i(c-d)) z_1 - (a+ib) \zbar_0 \\\label{eq:trafo4}
\zbar_1' &=& (1-i(c-d)) \zbar_1 - (a-ib) z_0 \,.
\eeqa
Here $a,b,c$ generate $SU(2)_L$-transformations ($c$ generates $U(1)_L$), while $d$ generates $U(1)_R$ rotations. By evaluating the transformation law for (operator-valued) one-forms 
\eq{A'_{i}(z')\extd z^{\prime i}  = A_{i}(z)\extd z^i}{eq:Ftransform}
to first order in the rotation parameters, one can show that the U(1) one-instanton solution \eqref{eq:oneinstantonA} leaves the full $SU(2)_L\times U(1)_R$ invariant. 

\subsection{Implications for the Dual Gauge Theory}

The dual field theory of a probe D7 brane embedded in $AdS_5 \times S^5$ as in \eqref{eq:masslessemb}  is a $\mathcal{N}=2$ supersymmetric $U(N_c)$ gauge theory, which has one massless hypermultiplet in the fundamental representation of the gauge group, coupled to the $\mathcal{N}=4$ vector multiplet in the adjoint representation.
As we have argued in section~\ref{sect:FI}, switching on a constant anti-selfdual B field on the gravity side is dual to FI terms for the auxiliary fields $(D, F_1, F_2)$ of the $\mathcal{N}=2$ $U(1) \subset U(N_c)$ vector multiplet. 

Due to the special form \eqref{eq:B-Field} of the B field (with $b_1=-b_2$), only the FI term for the auxiliary D field is present. The global symmetries\footnote{The global symmetries for a $\mathcal{N}=2$ supersymmetric $U(N_c)$ gauge theory with hypermultiplets in the fundamental representation are discussed in section \ref{sect:comminst}.} of the gauge theory, which coincide with the symmetries of gravity side, are given by
\begin{equation}
SO(3,1) \times SU(2)_{L} \times U(1)_{R} \times U(1)_z \times U(1)_F,
\end{equation}
with $\mathcal{R}$-Symmetry $U(1)_{R}$, which is the remaining
unbroken part 
of the $SU(2)_R$, and flavour symmetry $U(1)_F.$ 

The F- and D-term equations of this gauge theory  
\beqa \label{eq:DF1flavour}
0 &=& \Phi_3 q = \tilde{q} \Phi_3  \, ,  \label{F1}\\ 
0 &=& [\Phi_1,\Phi_3] = [\Phi_2,\Phi_3] \, ,\label{F2}\\
0 &=& q\tilde{q} + [\Phi_1,\Phi_2] \, , \label{F3}\\
\zeta \mathbbm{1}_{N_c \times N_c}
 &=& |q|^2-|\tilde{q}|^2+[\Phi_1,\Phi_1^\dagger]+[\Phi_2,\Phi_2^\dagger]\,
 \label{D1flavour} .
\eeqa
We will show at the end of the section, that these F- and D-term equations can still be satisfied simultaneously.

First let us consider the case in which all squark vevs $\tilde q$ and $q$ vanish. Since $U(1)$ factors drop out of the commutator terms in \eqref{D1flavour}, it is impossible to solve \eqref{D1flavour}. Therefore the pure Coulomb state is not supersymmetric as expected.

Now we are interested in the mixed Coulomb-Higgs states, which is dual to the
Nekrasov-Schwarz solution with instanton charge $k=1$. Since the number of nonvanishing 
squark components $q_a$ is related to the instanton charge, only one (colour) component of the squark fields $q$ and $\tilde q$ in the dual gauge theory are nonzero. These components will be called $q_1$ and $\tilde q_1$ respectively. Due to the ansatz \eqref{eq:coulombhiggs1} and \eqref{eq:coulombhiggs2} for the mixed Coulomb/Higgs branch equation \eqref{F1} is trivially satisfied. Using this ansatz the F-term equation \eqref{F2} gives the constraint \eqref{eq:FPhi12BZeroHiggs}. Furthermore the equations \eqref{F3} and \eqref{D1flavour} reduce to
\eq{0 = q_1\tilde{q_1} = |q_1|^2 - |\tilde{q}_1|^2 - \zeta\, .}{eq:k1Nf1}
The other D-term equations involving $q_a$ and $\tilde{q}_a$ for $a=2, \dots, N_c$ cannot be satisfied.

Solving \eqref{eq:k1Nf1}, we find that the gauge theory with only one flavour has a discrete state with squark vev at strong coupling. After an appropriate $U(1)_R$ rotation the nonvanishing squark vev is given by $q_1=\sqrt{\zeta}$, $\tilde{q}_1 = 0$. We recognize that the squark vev is in one-to-one correspondence with the moduli of the $U(1)$ noncommutative instanton on the gravity side. This is in agreement with the fact that except of the position moduli, noncommutative $U(1)$ instantons do not have additional moduli and their  size modulus is given by the noncommutativity parameter of the underlying spacetime. 

Due to the squark vevs $q_1=\sqrt{\zeta}$ and $\tilde{q}_1=0,$ the flavour symmetry $U(1)_F,$ the $\mathcal{R}$-symmetry $U(1)_R$ and the $U(1)$ part of the $U(N_c)$ gauge group are broken. Since the squark vevs $(q, \tilde{q})$ transform under $U(1)_{F}$ as
\begin{equation}
q \rightarrow e^{i \alpha_F} q, \quad \tilde q \rightarrow e^{-i \alpha_F} \tilde q
\end{equation}
the $U(1)_{F}$ rotation can be undone by an appropriate $U(1)_{R}$ transformation of the form
\begin{equation}
q \rightarrow e^{i \alpha_R} q, \quad \tilde q \rightarrow e^{i \alpha_R} \tilde q.
\end{equation}
Therefore the diagonal subgroup $\text{diag}(U(1)_{R} \times U(1)_F)$ is preserved by the squark vev. Note that this further breaking of the symmetries is not seen in the rather naive calculation of the previous section. We believe that the breakdown to $\text{diag}(U(1)_R\times U(1)_F)$ is due to the mixing of spacetime symmetries with gauge symmetries in noncommutative gauge theories \cite{Szabo:2001kg}, but as we do not know the action of a noncommutative gauge theory on the curved space $AdS_5\times S^3$, we have to leave the proof of this fact for future work. Intuitively, as in the AdS/CFT context global symmetries in the field theory do correspond to local symmetries on the gravity side, one would have to work out the noncommutative diffeomorphisms and gauge transformations which leave invariant the asymptotic behaviour $A_i \simeq |\vec{y}|^{-3}$ of the gauge potential \eqref{eq:oneinstantonA}. 

The symmetries, which are present in this state, are summarised in table~\ref{tab:D-Brane_symmetries}. The colour coding is as follows: Red, yellow and blue indicate the range of coordinates which are acted upon by the respective groups given there. The mixed-color orange and green fields indicate coordinates which are acted upon by both of the two adjacent symmetry transformations.

\TABLE[!ht]{
\setlength{\arrayrulewidth}{2pt}
\arrayrulecolor{white}
        \begin{tabular}{*{11}{c}}\hline
                                        \multicolumn{1}{>{\columncolor[gray]{.7}}c|}{Object} & \multicolumn{10}{
                                                >{\columncolor[gray]{.7}}c
                                        }
                                        {
                                                Coordinates
                                        }
                \\ \hline 
                &       \multicolumn{4}{
                                                >{\columncolor{lightviolet}}
                                                c|
                                        }
                                        {
                                                $x^{\mu}$
                                        }
                        &
                                \multicolumn{4}{
                                                >{\columncolor{lightviolet}}
                                                c|
                                        }
                                        { \begin{tabular}{cccc}
                                                \multicolumn{4}{c}{$y_m$} \\
                                                \multicolumn{1}{c}{\hspace{-2.7truecm}$\rho$} & \multicolumn{3}{c}{\hspace{2.0truecm}$S^3$}
                                                \end{tabular}
                                        }
                        &
                                \multicolumn{2}{
                                                >{\columncolor{lightviolet}}
                                                c
                                        }
                                        {
                                                $z^i$
                                        }
                \\\hline \multicolumn{1}{>{\columncolor[gray]{.8}}c|}{$AdS_5\times S^5$} & 
                \multicolumn{4}{>{\columncolor{lightred}}c}{
                                                $SO(4,2)$
                                }
                                &
                \multicolumn{1}{>{\columncolor{lightorange}}c}{} & 
                \multicolumn{3}{>{\columncolor{lightyellow}}c}{
                                                $SO(6)\simeq SU(4)$
                                }
                                &
                \multicolumn{2}{>{\columncolor{lightgreen}}c}{}
                \\
                \hline \multicolumn{1}{>{\columncolor[gray]{.8}}c|}{$D7$} &
                                \multicolumn{4}{>{\columncolor{lightred}}c}{
                                                $SO(4,2)$
                                }
                        &
                        \multicolumn{1}{>{\columncolor{lightorange}}c}{}
                        &
                        \multicolumn{3}{>{\columncolor{lightyellow}}c}{
                                                $SU(2)_L \times SU(2)_R$
                                }
                        &
                        \multicolumn{2}{>{\columncolor{lightblue}}c}{
                                                $U(1)_z$
                                }
                \\
                \hline \multicolumn{1}{>{\columncolor[gray]{.8}}c|}{$\ast B=-B$} &
                                \multicolumn{4}{>{\columncolor{lightred}}c}{
                                                $SO(4,2)$
                                }
                        &
                        \multicolumn{1}{>{\columncolor{lightorange}}c}{}
                        &
                        \multicolumn{3}{>{\columncolor{lightyellow}}c}{
                                                $SU(2)_L \times U(1)_R$
                                }
                        &
                        \multicolumn{2}{>{\columncolor{lightblue}}c}{
                                                $U(1)_z$
                                }
                \\
\hline \multicolumn{1}{>{\columncolor[gray]{.8}}c|}{Eq.~\eqref{eq:oneinstantonA}} &
                                \multicolumn{4}{>{\columncolor{lightred}}c}{
                                                $SO(4,2)$
                                }
                        &
                        \multicolumn{1}{>{\columncolor{lightorange}}c}{}
                        &
                        \multicolumn{3}{>{\columncolor{lightyellow}}c}{
                                                $SU(2)_L \times \text{diag}(U(1)_R\times U(1)_F)$
                                }
                        &
                        \multicolumn{2}{>{\columncolor{lightblue}}c}{
                                                $U(1)_z$
                                }
        \end{tabular}
\caption{The symmetry breaking pattern.}
\label{tab:D-Brane_symmetries}
}

Finally for completeness 
let us consider the case of the Higgs vacuum, for which all D3 branes
are dissolved in the D7, and for which an AdS/CFT dual is not describable in the
probe limit. 
The Higgs vacuum is given by $\Phi_3=0$ and $|q_a|=\sqrt{\zeta}, 
\tilde{q}_a=0$ for $a=1, \dots, N_c.$ Furthermore all commutators 
involving $\Phi^1,\Phi^2$ and $\Phi^3$ vanish. It is easy to verify 
that the F- and D-term equations are simultaneously satisfied. The Higgs 
vacuum is therefore supersymmetric. This vacuum corresponds to a charge 
$N_c$ noncommutative $U(1)$ instanton.

    \section{Conclusions}
    \label{sect:conclusions}

In the construction presented, the mixed Coulomb-Higgs branch of the
field theory with $N_f$ flavours with FI term for $k$ colour directions is
dual to a probe of $N_f$ D7 branes with charge $k$ non-commutative
instanton in $AdS_5 \times S^5$. The non-commutativity is generated by an
anti-selfdual B field in the internal directions on the D7 probe.

A central result is that the $U(1)$ factor of the $U(N)$ gauge group
in the boundary field theory plays an important role, though it is broken
eventually for a special solution to the F- and D-term equations dual to the
Nekrasov-Schwarz instanton.  Our main
evidence for the presence of the FI term is its matching with the
quantum numbers of the B field in the supergravity background. It will be
interesting to study the singleton mechanism involved
further.

For the present analysis, we have worked in an adiabatic probe approximation
where $N_c-k$ D3 branes generate the $AdS_5 \times S^5$ geometry, while
$k$ D3 branes are dissolved as instantons
in the D7 brane probe. Evidently, $k$ must be small. This
approximation is suited for explaining the general mechanism of the
duality in presence of a non-commutative instanton. As both supersymmetry and gauge symmetry
is broken in this setup, an interesting application of the proposal made in this work would be to analyze
the corresponding Higgs mechanism and find the Goldstino in the spectrum of the $D7$ brane fluctuations.

For the future, it will be interesting to investigate a similar duality for the more
involved case where the scenario is stabilized, for instance by
considering D7 brane probes in a singular geometry. This may allow for
constructing  gravity dual descriptions of metastable vacua. It will also be 
interesting to study supersymmetry breaking for D7 branes in the warped throat 
geometry, for instance based on the results of \cite{Krause:2007jk,Baumann:2007ah}. 
Both issues - i.e~the singleton and the stabilisation mechanism - may
also be addressed by considering D7 brane probes in the recently
discussed deformed Sasaki-Einstein geometries of \cite{Martelli:2008cm}. 

        \subsection*{Acknowledgements}
We are grateful to Robert Helling, Andreas Karch, Paul Koerber, Olaf Lechtenfeld, Alberto Lerda, Yi Liao, Luca Martucci, Jeong-Hyuck  Park, Mukund Rangamani, Dima Vassilevich, Robert Wimmer and Marco Zagermann for useful discussions. This work was supported in part by the Cluster of Excellence "Origin and Structure of the Universe". J.E. acknowledges hospitality of the Institute for Nuclear Theory,  University of Washington, Seattle, during the final stages of this work.

\appendix

        \section{Expansion of the Determinant in the DBI-Action}
        \label{sect:determinant_expansion}
For completeness, we write down the most important steps of the expansion of the $D7$-brane action (\ref{eq:effective_action_FI}) to second order $2\pi\alpha'$ in more detail. In the following discussion we abbreviate\footnote{In this appendix we suppress the colour traces to avoid confusion between them and traces over euclidean space indices.}
\begin{equation}
E_{ab}=\mathcal{P}_{ab}[g_{\mu\nu} + B_{\mu\nu}].
\end{equation}
Using the expansion
\begin{equation}
\label{eq:determinant_expansion}
\begin{split}
&\sqrt{-\det(E_{ab} + 2 \pi \alpha' F_{ab})}   = \sqrt{-\det E_{ab} \det(\delta_{ab} + 2 \pi \alpha' (E^{-1} F)_{ab})}\\
&= \sqrt{-\det E_{ab}} \cdot \left( 1 + \frac {2 \pi \alpha'}
 2 {\rm Tr} \, E^{-1}  F  + \frac {\left(2 \pi \alpha'\right)^2}
 8 \left( {\rm Tr} \, E^{-1} F \right) ^2  \right. \\ & \hspace{4cm}
\left. - \frac
 {\left(2 \pi \alpha'\right)^2} 4 {\rm Tr} \left( E^{-1} F \right)^2 +
 \dots \right),
\end{split}
\end{equation}
the action of the $D7$-brane up to order $\alpha^{\prime 2}$ 
is given by
\begin{eqnarray}
S_{D7}&=&S^{(0)} + 2 \pi\alpha^\prime S^{(1)} + \left( 2 \pi \alpha'
\right)^2 S^{(2)} + \mathcal{O}(\alpha^{\prime \, 3})\, ,  \\
S^{(0)}&=& - T_7 \int\limits_{D7} dx^\mu dy^m \sqrt{-\det E_{ab}} \pm
\frac{T_7}{2} \int\limits_{D7} C_{(4)} \, \wedge B \wedge B  \, , 
\eeqa
\beqa
S^{(1)}&=& - \frac{T_7} 2 \int\limits_{D7} dx^\mu dy^m \sqrt{-\det
  E_{ab}} E^{ab} F_{ba} \pm T_7 \int\limits_{D7} C_{(4)} \wedge  B
\wedge F \, , \\
S^{(2)}&=& - T_7 \int\limits_{D7} dx^\mu dy^m \sqrt{-\det E_{ab}} \left[ \frac 1 8 \left( E^{ab} F_{ba} \right)^2 - \frac 1 4 E^{ab}F_{bc}E^{cd}F_{da} \right] \\
& &\, \pm \frac {T_7}{2} \int\limits_{D7} C_{(4)} \wedge F \wedge F \, .
\end{eqnarray}
The inverse matrix of $E_{ab}$ will be denoted by upper indices,
i.e. $E^{ab}$, with
\begin{equation}
E^{ab} E_{bc} = \delta^a_c.
\end{equation}
Switching on the selfdual, constant $B$ field (\ref{eq:B-Field}), a non-vanishing $U(1)$ field strength in the directions transversal to the $D3$-brane, and using the massless embedding of the $D7$-brane, we obtain
\begin{eqnarray*}
\sqrt{-\det (E_{ab})}&=&\frac{H_3(r)+b^2}{H_3(r)} = 1 +
\frac{b^2}{H_3(r)} \, , \\
E^{ab} F_{ba} &=& 2 \, b \, A \left( F_{45} + F_{67} \right) = \frac{2
  b}{H_3(r)+b^2} \left( F_{45} + F_{67} \right) \, , \\
\frac 1 8 \left( E^{ab} F_{ba} \right)^2 - \frac 1 4
E^{ab}F_{bc}E^{cd}F_{da} &=& \frac 1 {4 (H_3(r)+b^2)^2} \left( H_3(r)
  F_{ab} F_{ab} + \frac{b^2} 2
 \epsilon_{klmn} F_{kl} F_{mn} \right) \, , \\
C_{(4)} \wedge B \wedge B &=& \frac{2 b^2}{H_3(r)} d^4 x \, d^4 y \, ,
\\
C_{(4)} \wedge B \wedge F &=& \frac b {H_3(r)} \left(F_{45} + F_{67}
\right) d^4x d^4y \, ,
\\
C_{(4)} \wedge F \wedge F &=& \frac 1 {4 \, H_3(r)} \epsilon_{klmn}
F_{kl} F_{mn} d^4x d^4y \, ,
\end{eqnarray*}
where $d^4x \, d^4y = dx^0 \wedge \dots \wedge dx^3 \wedge dy^4 \wedge \dots \wedge dy^7.$
Using these results, we have
for the $D7$-brane action $S_{D7}=S_{DBI} + S_{WZ}$ up to order
$\alpha^{\prime \, 2}$
\begin{eqnarray}
S_{D7}  &=&S^{(0)} + 2 \pi\alpha^\prime S^{(1)} + \left( 2 \pi \alpha'
\right)^2 S^{(2)} + \mathcal{O}(\alpha^{\prime \, 3})  \, , \\
S^{(0)} &=& - T_7 \int dx^\mu \, dy^m \, 1 \, , \\
S^{(1)} &=& 0 \, , \\
S^{(2)} &=& - \frac{T_7}{4} \int  dx^\mu \, dy^m \, \frac{1}{H_3(y) +
  b^2} \left( F_{ab} F_{ab} - \frac 1 2 \epsilon_{klmn} F_{kl} F_{mn}
\right)  \\
&=& -  \frac{T_7}{2} \int  dx^\mu \, dy^m \, \frac{1}{H_3(y) + b^2}
F_- F_- \, ,
\end{eqnarray}
which is precisely the action (\ref{eq:effective_action_FI}).

\end{document}